\documentclass{PoS}
\graphicspath{{./Figures/}}                           

\newcommand{\eg}{\textit{\mbox{e.g.\ }}}              
\renewcommand{\Re}{\mathfrak{Re}\,}                   
\newcommand{\Tr}{\mbox{Tr}}                           
\newcommand{\Fig}[1]{{Fig.~\ref{#1}}}

\newcommand{\Eq}[1]{Eq.~(\ref{#1})}

\newcommand{\QCDSF}{\textsf{QCDSF}}

\newcommand{\preprint}{\newline%
  \begin{picture}(0,0)
  \put(293,110){\rm\small HU--EP--06/34, LU--ITP 2006/018}
  \end{picture}}

\title{Lattice study of the infrared behavior of QCD Green's functions
in Landau gauge\thanks{This work was supported by the DFG under the
  contract FOR 465.}\preprint} 

\ShortTitle{Lattice study of QCD Green's functions}

\author{A.~Sternbeck\thanks{Speaker. Supported by the DFG-funded
    graduate school GK~271. Address until August 31, 2006:\newline
    Humboldt-Universit\"at zu Berlin, Institut f\"ur Physik,
  D-12489 Berlin, Germany.}\\ 
  Centre for the Subatomic Structure of Matter (CSSM), School of Chemistry  
  \& Physics,
  University of Adelaide, SA 5005, Australia\\
  E-mail: \email{andre.sternbeck@adelaide.edu.de}
}

\author{E.--M.~Ilgenfritz and M.~M\"uller-Preussker\\
  Humboldt-Universit\"at zu Berlin, Institut f\"ur Physik,
  12489 Berlin, Germany\\
  E-mail: \email{ilgenfri@physik.hu-berlin.de}, \email{mmp@physik.hu-berlin.de}
}
\author{A.~Schiller\\
  Universit\"at Leipzig, Institut f\"ur Theoretische 
  Physik, 04109 Leipzig, Germany\\
  E-mail: \email{Arwed.Schiller@itp.uni-leipzig.de}
}
\author{I.~L.~Bogolubsky\\
  Joint Institute for Nuclear Research,
  141980 Dubna, Russia\\
  E-mail: \email{bogolubs@jinr.ru}
}

\abstract{We summarize the current status of our numerical results
  for the gluon and ghost propagators and for the Kugo-Ojima
  confinement parameter in quenched $SU(3)$ lattice Landau gauge
  theory. The data for the propagators are compared to our
  results obtained in the case of full QCD, simulated using two flavors of
  dynamical clover-improved Wilson fermions. We demonstrate that the infrared
  behavior of the ghost propagator is consistent with the Kugo-Ojima
  confinement criterion. Explicit violation of reflection
  positivity by the gluon propagator is shown. Additionally, we
  present results of a running coupling constant both at low and at
  large momenta.}

\FullConference{XXIV International Symposium on Lattice Field Theory\\
		 July 23-28 2006\\
		 Tucson, Arizona, US}

\begin{document}

\section{Introduction}

In the last years progress has been made to improve our understanding of
QCD Green's functions in different gauges. This is important, because it
might allow us someday to describe all hadronic features in the
continuum directly in terms of those functions. A deeper knowledge of
QCD Green's functions would help us also to understand the basics of  
nonperturbative phenomena like gluon and quark confinement or
dynamical chiral symmetry breaking from first principles. For example,
the realization of the Kugo-Ojima confinement scenario
\cite{Kugo:1979gm} in QCD in covariant gauges is encoded
in the infrared behavior of the gluon and ghost 2-point
functions. Therefore, the investigation of QCD Green's functions is not
only of interest for a coherent description of hadronic states but
also for an understanding of confinement. 

At large (Euclidean) momenta QCD Green's functions can be described
in terms of perturbation theory. However, at intermediate and low momenta
nonperturbative methods are indispensable to arrive at a complete picture.
There are different approaches to deal with nonperturbative QCD, but
they all have their own limitations. Therefore, a comparison of
results is important. For example, studies of truncated
Dyson-Schwinger equations (DSEs) for the gluon and ghost propagators in Landau
gauge came to the conclusion that at very low momenta these propagators are
governed by power laws with interrelated exponents which then results
in a coupling constant running to a non-trivial infrared fixed point
\cite{vonSmekal:1997isvonSmekal:1998yu}. This 
infrared behavior has been confirmed independently by studies of exact
renormalization group equations \cite{Pawlowski:2003hq} and
investigations based on the Fokker-Planck-type diffusion equation of
stochastic quantization \cite{Zwanziger:2001kw}. At intermediate
momenta, however, predictions \eg based on DSE studies are uncertain
due to the truncations involved, but in this region lattice Monte
Carlo (MC) simulations can help. 

\section{Lattice results for the gluon and ghost propagators}

We have performed MC simulations in the quenched
approximation of $SU(3)$ lattice gauge theory 
at four values of the coupling constant ($\beta=5.7$, 5.8, 6.0, 6.2)
using the standard gluonic Wilson action. Thereby, the lattice size
ranged between $16^4$ and $56^4$. To study also the influence of
fermions we have analyzed gauge field configurations 
generated with $N_f=2$ dynamical flavors of clover-improved Wilson
fermions using the same gauge action\footnote{Those 
configurations were provided to us by the QCDSF collaboration.}. 
All gauge configurations have been fixed to Landau gauge by maximizing
the Landau gauge functional
\begin{equation}
  F_{U}[g] = \frac{1}{4V}\sum_{x}\sum_{\mu=1}^{4}\Re\Tr
  \;{}^{g}U_{x,\mu} \qquad\textrm{with}\quad  {}^{g}U_{x,\mu}=g_x\,
  U_{x,\mu}\,g^{\dagger}_{x+\hat{\mu}}
  \label{eq:functional}
\end{equation}
using either over-relaxation or Fourier-accelerated
gauge-fixing. Of course, the maxima are not unique, but we will 
neglect the influence of Gribov copies in the following. The reader is
referred to Ref.~\cite{Sternbeck:2005tk} for a more systematic account
of the Gribov ambiguity in this context. 

\smallskip

After gauge-fixing we have calculated the gluon and ghost propagators
on all (gauge-fixed) configurations. The definition of both Green's
functions and their dressing functions is standard and our notation can be 
found in detail in Ref.~\cite{Sternbeck:2006rd}. To reduce finite
volume and discretization effects we have applied cone and
cylinder cuts to our data \cite{Leinweber:1998uu}. 

\begin{figure}[b]
  \centering
\begin{minipage}[t]{0.48\textwidth}
  \includegraphics[height=5.2cm]{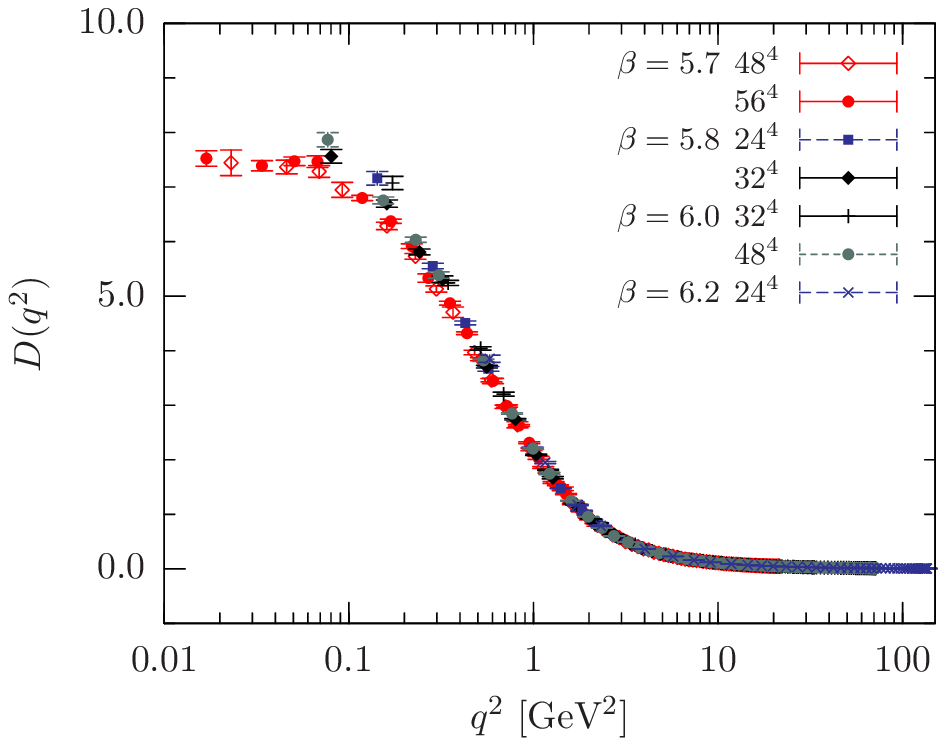}
  \label{fig:gl_qq_fc}
 \caption{The (quenched) gluon propagator $D$ as a function of $q^2$.}
\end{minipage} 
\hfill
\begin{minipage}[t]{0.48\textwidth}
  \includegraphics[height=5.1cm]{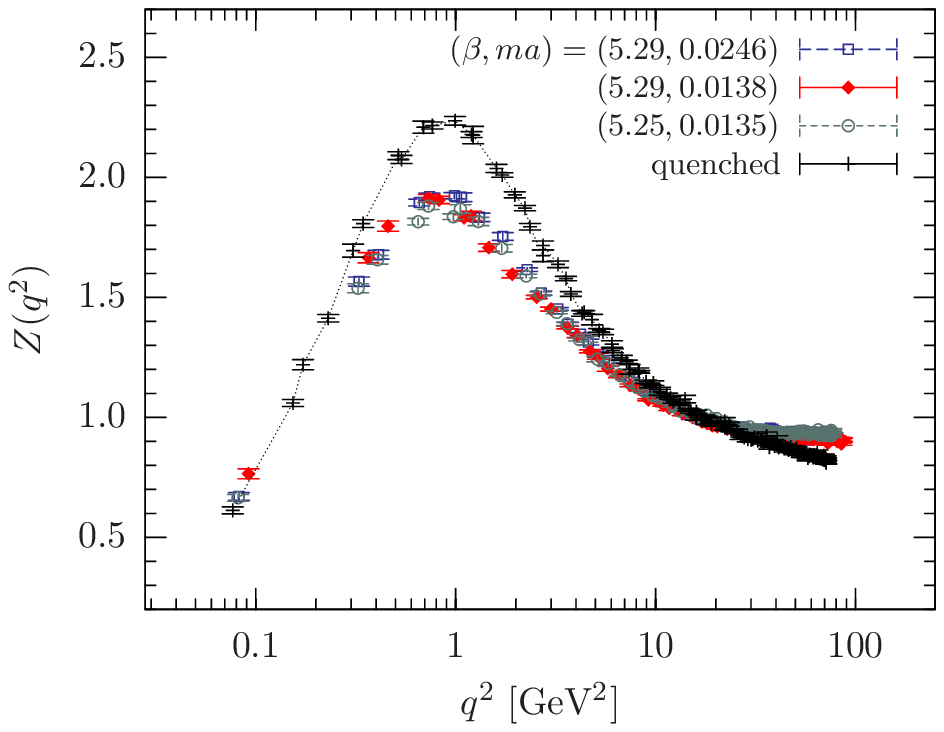}
  \caption{The gluon dressing function $Z$ for the quenched and unquenched
  case as a function of $q^2$}
\label{fig:gl_dress_qq_dyn}
\end{minipage}
\end{figure}

\begin{floatingfigure}[r]
  \centering
  \includegraphics[width=0.45\textwidth]{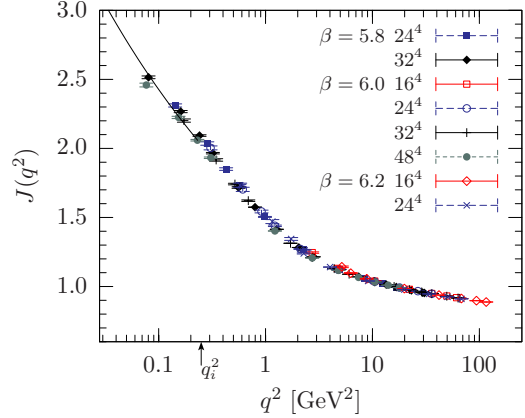}
  \caption{The ghost dressing function $J$ for the quenched case as a
    function of the momentum $q^2$. The line represents a fit of a
    power law to the data at $q^2<q^2_c$ resulting in the exponent
    $\kappa\approx0.2$.} 
\label{fig:gh_dress_qq_fc}
\end{floatingfigure}
Considering first the quenched case, in 
\Fig{fig:gh_dress_qq_fc} we show data for the ghost dressing function
$J(q^2)$ (renormalized at $\mu=4~\textrm{GeV}$) as a function of the
momentum $q^2$. As expected the ghost dressing function seems to
diverge at vanishing momentum. We have tried to fit the infrared
power law as expected from DSE studies to our data. However, within
the region of lower momenta a power law does not 
describe the data that well. The ghost dressing function seems
to increase logarithmically as the momenta become small. In any case,
the infrared exponent extracted from the fit, $\kappa\approx0.2$, is
much smaller then it is expected to be, namely $\kappa\approx0.59$
\cite{Zwanziger:2001kw,Lerche:2002ep}. With respect to our data for the
gluon propagator, shown in \Fig{fig:gl_qq_fc}, 
we think that we are still not in a region of momenta where the
mentioned infrared power law can be verified. For this the gluon
propagator $D$ has to vanish at zero momentum which cannot be concluded
from our current data (see \Fig{fig:gl_qq_fc}). To what extent finite
volume effects can be blamed for such a (different) behavior found on
the lattice needs to be clarified in future yet.  

To improve our understanding of lattice Landau gauge theory we have
also studied  the influence of (clover-improved) Wilson fermions on
the ghost and gluon propagators. We find that the ghost propagator
stays almost unchanged if fermions are added to the gauge action. 
In contrast to this, fermions affect the gluon propagator at large and
intermediate momenta, in particular where the gluon propagator exposes
its characteristic enhancement compared to the free propagator. To
illustrate this, in \Fig{fig:gl_dress_qq_dyn} we show our data for the
gluon dressing function obtained for quenched and full QCD.  

\smallskip
As a part of the project we also performed simulations at
$\beta=6.0$ using large asymmetric lattice sizes, namely $16^3\times128$ and
$24^3\times128$. By comparing the data obtained on these lattices for
the ghost and gluon propagators to our results using
symmetric lattices (\eg $48^4$) we found large systematic
effects at low momentum due to the asymmetry involved,
in particular, for the lowest on-axis momenta along the elongated
'time' direction.\footnote{For more details about this but also other systematic
effects have a look at the Ph.D. thesis \cite{Sternbeck:2006rd} of one of
us.} It is not excluded that sensible results at lower momenta for
the gluon and ghost propagators can be extracted from asymmetric
lattices by careful extrapolations. Attempts in that directions are
made \eg in \cite{Silva:2005hbSilva:2006bs}.

\begin{figure}[t]
  \centering
\begin{minipage}[t]{0.48\textwidth}
  \includegraphics[height=6.3cm]{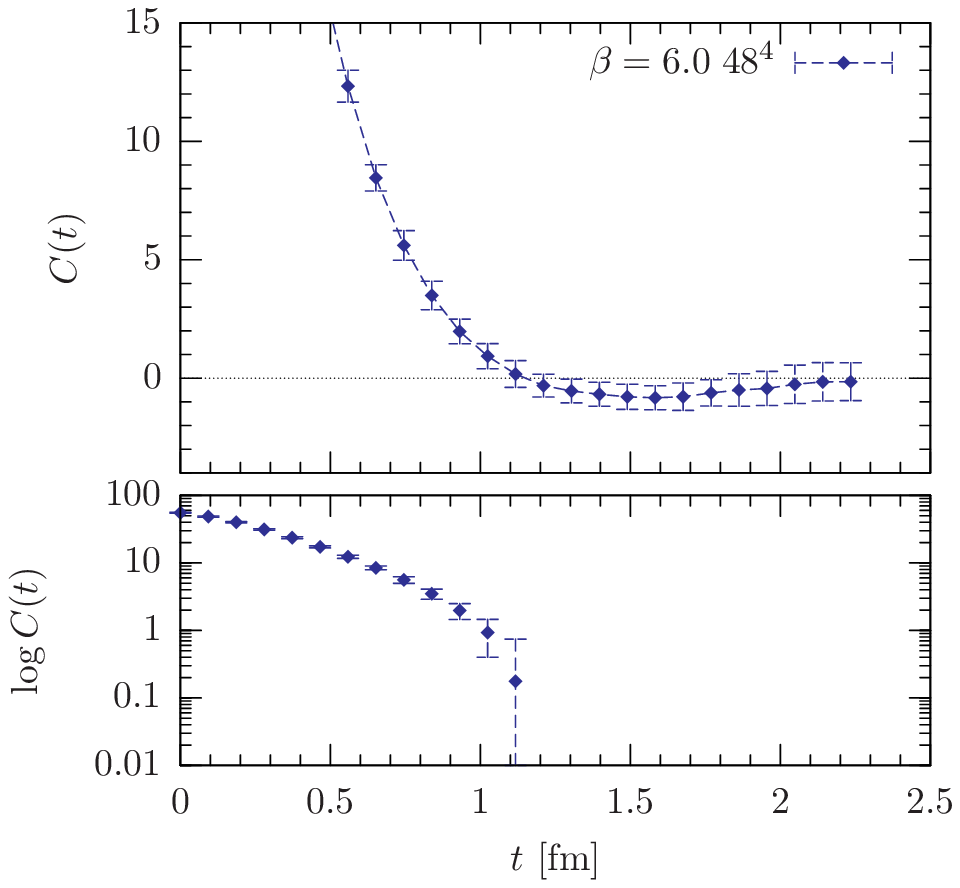}\vspace{-1ex}
  \caption{The upper panel shows the zero three-momentum propagator $C(t)$
    of the gluon fields in quenched QCD at $\beta=6.0$ for a $48^4$
    lattice as a function of time. In the lower panel the 
    same data are shown, however, as $\log C(t)$ for $C(t)>0$.}
\label{fig:c_t}
\end{minipage} 
\hfill
\begin{minipage}[t]{0.48\textwidth}
  \includegraphics[height=6.3cm]{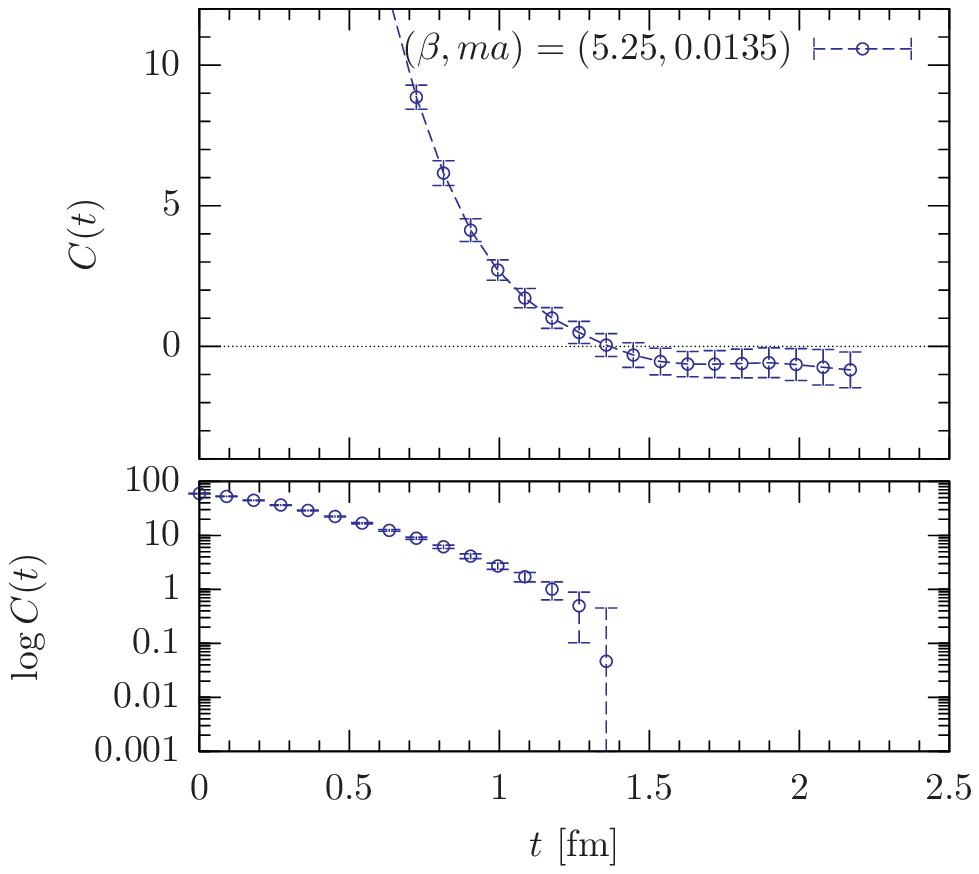}\vspace{-1ex}
  \caption{The same as in the left figure, however, for the 
    gluon propagator in full QCD on a $24^3\times48$ lattice for one particular
    mass $ma$.}
\label{fig:c_t_dyn}
\end{minipage}
\end{figure}

\section{Confinement criterion I: Violation of reflection positivity}

In view of the general objectives of this project it is also important to check
whether our data for the propagators satisfy necessary
criteria for confinement. For example, the gluon propagator has to 
violate reflection positivity, because otherwise gluons could be interpreted
in terms of stable particle states.\footnote{Note that the ghost
  propagator violates reflection positivity trivially and hence ghosts
  are explicitly unphysical.}
To check whether reflection positivity is violated by the gluon propagator 
we have calculated the space-time (lattice) correlator at zero spatial
momentum as function of time
\begin{equation}
 \label{eq:C_t_latt}
  C(t) := \frac{1}{\sqrt{L_T}}\sum_{k_4=0}^{L_T-1}
  D(\vec{0},k_4) \exp\left\{ \frac{2\pi ik_4t}{L_T}\right\} 
\end{equation}
using our data for the gluon propagator $D(\vec{0},k_4)$. The result
for the quenched case ($\beta=6.0$, $48^4$ lattice) is shown in
\Fig{fig:c_t}. Obviously, reflection positivity is violated in a
finite range of~$t$. The same holds in the unquenched case as can be seen
in \Fig{fig:c_t_dyn}, even though the lattice size in this case is
too small such that the data do not bend over towards zero at
larger~$t$ as in the quenched case (\Fig{fig:c_t}). Note that we have
seen this happening also in our quenched data at $\beta=6.0$ using a $32^4$
lattice (not shown).

\section{Confinement criteria II: The Kugo-Ojima confinement parameter}

Another criterion for confinement for QCD in covariant gauges was
given by Kugo and Ojima long time ago \cite{Kugo:1979gm}. According to
their scenario colored asymptotic states, if any, cannot be detected
in the physical subspace due to the quartet mechanism. This is
realized if a function $u(p^2)$, defined as \cite{Kugo:1995km,Alkofer:2000wg}
\begin{equation}
  \label{eq:def_u}
  \int d^4x\; e^{ip(x-y)} \left\langle D^{ae}_{\mu} c^e(x) g f^{bcd}
  A^d_{\nu}(y) \bar{c}^{c}(y)\right\rangle =: \left( \delta^{\mu\nu} -
    \frac{p_{\mu}p_{\nu}}{p^2}\right) \delta^{ab}\,u(p^2),
\end{equation}
\begin{floatingfigure}
  \centering
  \includegraphics[width=0.45\textwidth]{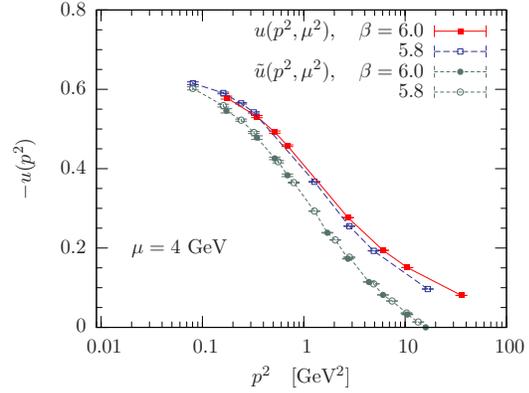}\vspace{-0.3cm}
   \caption{Data of $u(p^2,\mu^2)$ at $\beta=5.8$ and 6.0 are shown
     using squares. Additionally, data of
     $\tilde{u}$ are shown at the same $\beta$ values
     (circles). All data refer to the same quenched 
     configurations on a $32^4$ lattice and are renormalized at
     $\mu=4~\textrm{GeV}$ as described in
    \cite{Sternbeck:2006rd}. Lines are drawn to guide the eye.} 
  \label{fig:ku_log_qq}
\end{floatingfigure}
{\noindent has the zero-momentum limit\vspace{-1ex}}
\begin{equation}
 \label{eq:u_lim_0}
  \textsf{u}:=\lim_{p^2\rightarrow 0} u(p^2) = -1 \;.
\end{equation}
In Landau gauge this limit is connected to an infrared
diverging ghost dressing function $J$ through \cite{Kugo:1995km}
\begin{equation}
  \label{eq:ghost_ku_q}
  J(p^2) = \frac{1}{1+u(p^2)+ p^2v(p^2)}\;
  \stackrel{p^2\rightarrow 0}{\longrightarrow}\; \frac{1}{1+\textsf{u}}\;\;.
\end{equation}
Here $v(p^2)$ is an unknown function of $p^2$ (see
\cite{Kugo:1995km} for a 
definition). We have made an attempt to confirm the realization of the
limit in \Eq{eq:u_lim_0}, not only by giving numerical evidence for a
diverging ghost dressing function (see above), but also by estimating
$u(p^2)$ itself at different momenta $p^2$ in our lattice
simulations\footnote{For details on our estimation and renormalization
  of $u(p^2)$ on the lattice we refer to
  Ref.~\cite{Sternbeck:2006rd}.}. Our estimates 
of $u(p^2)$, renormalized at $\mu=4\;\textrm{GeV}$, are shown in
\Fig{fig:ku_log_qq} as a function of $p^2$. There we also show data of
the ghost dressing function (renormalized at the same $\mu$) in the form
of an asymptote $\tilde{u}$ defined as
\begin{equation}
  \label{eq:u_asymptotic}
  \tilde{u}(p^2,\mu^2) := \frac{1}{J(p^2,\mu^2)} - 1\;.
\end{equation}
Obviously, $\tilde{u}(p^2)$ and $u(p^2)$ are different at finite
momentum, but according to \Eq{eq:ghost_ku_q}~
$u(p^2)$ has to reach asymptotically $\tilde{u}(p^2)$ in the
zero momentum limit. From our data shown in \Fig{fig:ku_log_qq} we can
confirm that the difference $|u(p^2)-\tilde{u}(p^2)|$ diminishes
with decreasing 
momentum, even though an extrapolation of the given data to vanishing
momentum is difficult to perform. For this, the explicit momentum
dependence of $u$ has to be known. Since $u(p^2)$ seems to
continuously approach $\tilde{u}(p^2)$ with decreasing momentum and, 
by definition, $\tilde{u}(p^2)$ is minus one at vanishing momentum for
a diverging ghost dressing function our studied momentum range of $u$
at least does not exclude the expected zero momentum limit (see
\Eq{eq:u_lim_0}).

\section{The running coupling constant}

Given the dressing functions of gluon and ghost propagators a
running coupling constant 
\begin{equation}
 \label{eq:running_coup2}
  \alpha_s(q^2) = \alpha_s(\mu^2)\,Z(q^2,\mu^2)\,J^2(q^2,\mu^2)
\end{equation}
\begin{floatingfigure}
\includegraphics[width=0.45\textwidth]{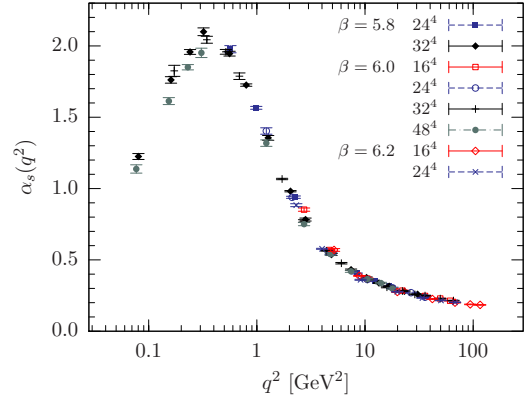}  
\label{fig:alpha_qq_fc}
\caption{The running coupling $\alpha_s(q^2)$ as a function of the 
    momentum $q^2$.} 
\end{floatingfigure}
\noindent based on the ghost-gluon-vertex can be calculated.
This definition relies on the assumption that in Landau gauge this vertex 
stays bare also beyond perturbation theory. We have given numerical
evidence in quenched and unquenched $SU(3)$ lattice gauge theory
confirming that the vertex renormalization constant
$\widetilde{Z}_1\approx1$ in a (asymmetric) $\textsf{MOM}$ scheme 
where the gluon momentum equals zero (see
\cite{Sternbeck:2006rd,Ilgenfritz:2006he}). Similar results indicating
this, directly and indirectly, were presented in lattice studies of
quenched $SU(2)$ gauge theory \cite{Cucchieri:2004sqBloch:2003sk}, but
also within the DSE approach \cite{Schleifenbaum:2004id}. 
Thus the definition of $\alpha_s(q^2)$ according to
\Eq{eq:running_coup2} is valid at least in that scheme.

In \Fig{fig:alpha_qq_fc} we show our current data of $\alpha_s$ in
the quenched case. One clearly sees it increasing towards
lower momenta as long as \mbox{$q^2>0.3$~GeV$^2$}. After passing a maximum 
at \mbox{$q^2\approx0.3$~GeV$^2$} it decreases. We have found the same
behavior on our sets of dynamical gauge configurations (see \eg Fig.~4 in
Ref.~\cite{Ilgenfritz:2006he}). Therefore, on the basis of the present data we
cannot confirm $\alpha_s(q^2)$ to approach a non-trivial infrared
fix-point, even less monotonously from below as expected from DSE studies. The
reason for this is still unclear, however, in the light of the DSE
results on a torus this might be a finite volume effect which we are
unable to resolve at the present stage.

\begin{figure}[b]
  \centering
  \includegraphics[width=0.44\textwidth]{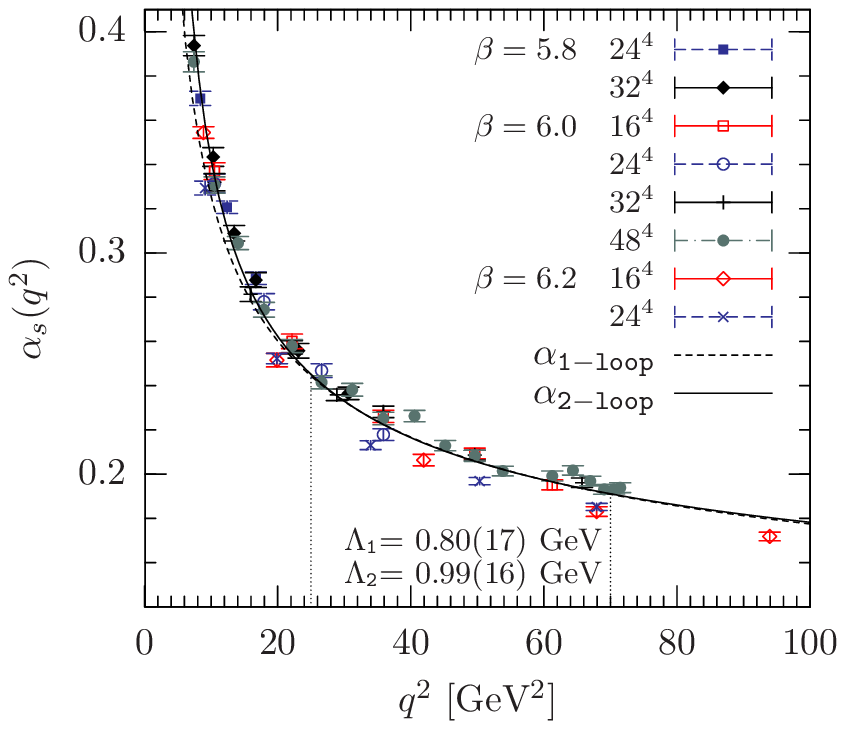}
  \quad
  \includegraphics[width=0.44\textwidth]{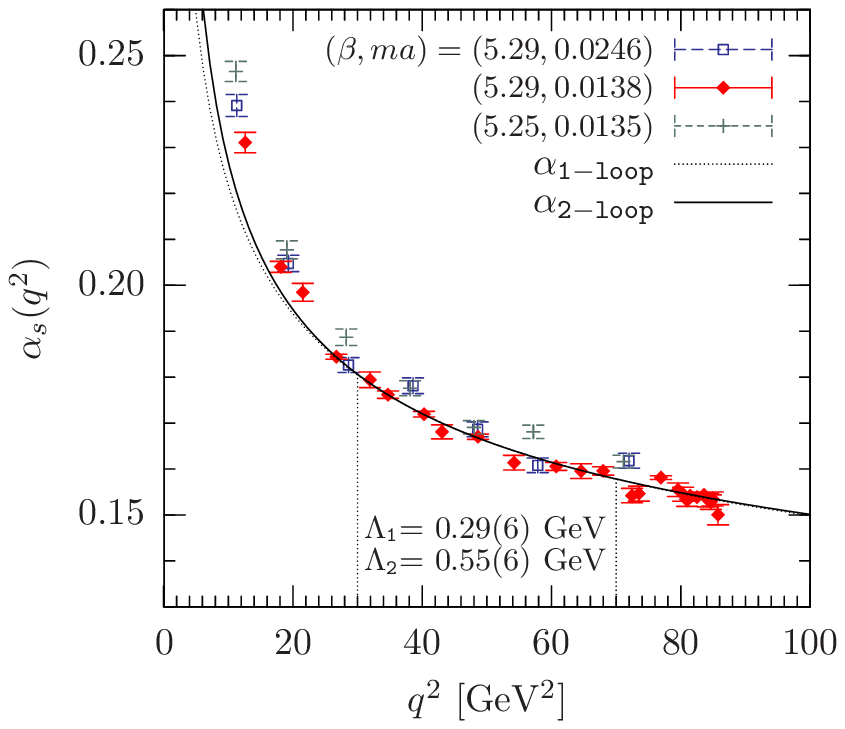}
  \caption{The running coupling $\alpha_s$ at large momenta for the
    quenched (left) and unquenched case (right). Since 
    discretization errors are quite large we used only
    data from the $48^4$ ($24^3\times48$) 
    lattice at $\beta=6.0$ ($\beta=5.29$, $ma=0.0138$) 
    to fit the 1-loop and 2-loop expressions of $\alpha_s$ to the data
    on the left (right) hand side of this figure.}  
  \label{fig:alpha_s_pert}
\end{figure}

Apart from the low-momentum region it is also interesting to look at
$\alpha_s(q^2)$ at larger momenta and to fit the corresponding 1-loop and 2-loop
expressions to the data. This is shown in \Fig{fig:alpha_s_pert} for
the quenched (l.h.s.) and unquenched case (r.h.s.). Obviously, the
data follow the one and two-loop expression in the marked interval.
However, large discretization errors are visible. Therefore,
simulations at smaller lattice spacings are worthwhile in order to get
more reliable values of $\Lambda$.  

\section{Conclusions}
\vspace{-1.5ex}
We have studied different aspects of lattice Landau gauge theory using
MC simulations of quenched and full QCD. Thereby, we have mainly
focused on the momentum dependence of the gluon and ghost
propagators. We have demonstrated here for the first time that our
lattice results for both propagators are consistent with different criteria for
confinement, even though we could not confirm the infrared behavior as
anticipated from DSE studies. Whether the gluon propagator at zero
momentum vanishes in the limit of infinite volume or not has to be clarified
in future studies. 

We could demonstrate that (dynamical) clover-improved Wilson
fermions affect the gluon but only negligibly the ghost
propagator. Importantly, the influence of fermions on the infrared
behavior of both propagators seems to be small. 

On the basis of our present data for $\alpha_s$ we cannot confirm it
to approach a non-trivial 
infrared fix-point monotonously from below. In any case, our data at
larger momenta agree with the 1-loop and 2-loop expressions of
$\alpha_s$, even though simulations at larger $\beta$ are
worthwhile to perform to suppress discretization effects and to
obtain more reliable results for $\Lambda$.

\bigskip
{\small
All simulations were performed on the IBM pSeries 690 at HLRN and on 
the MVS-15000BM at the Joint Supercomputer Center (JSCC) in Moscow.
This work was supported by the DFG under the contract FOR 465
(Forschergruppe Lattice Hadron Phenomenology), by the DFG-funded graduate
school GK~271 and with joint grants DFG~436~RUS~113/866/0 and
RFBR~06-02-04014. We thank the $\QCDSF$ 
collaboration for providing us their unquenched configurations which
we could access in the framework of the I3 Hadron-Physics initiative (EU
contract RII3-CT-2004-506078). We are grateful to Hinnerk St\"uben
for contributing parts of the program code. A.~St. acknowledges
discussions with Lorenz von Smekal.


\providecommand{\href}[2]{#2}\begingroup\raggedright\endgroup

\end{document}